\documentclass[aps,pra,twocolumn,amsmath,amssymb]{revtex4-1}
\usepackage{graphicx}
\usepackage{dcolumn}
\usepackage{bm}
\usepackage{color}
\usepackage{multirow}

\begin{document}
\author{Weibin Li}
\author{Lama Hamadeh}
\author{Igor Lesanovsky}
\affiliation{School of Physics and Astronomy, The University of Nottingham, Nottingham, NG7 2RD}

\title{Probing the interaction between Rydberg-dressed atoms through interference}
\date{\today}
\begin{abstract}
We study the dynamics of an atomic Bose-Einstein condensate in an optical lattice in which the electronic groundstate of each atom is weakly coupled to a highly excited Rydberg state by a far off-resonant laser. This dressing induces a switchable effective soft-core interaction between groundstate atoms which, in the lattice, gives rise to on-site as well as long-range interaction terms. Upon switching on the dressing laser the Bose-Einstein condensate undergoes a nontrivial collapse and revival dynamics which can be observed in the interference pattern that is created after a release of the atoms from the optical lattice. This interference signal strongly depends on the strength and the duration of the dressing laser pulse and can be used to probe and characterize the effective interaction between Rydberg-dressed atoms.
\end{abstract}
\maketitle
\section{Introduction}
Ultracold atoms trapped in optical lattices provide a highly versatile toolbox for the exploration of the statics and dynamics of many-body quantum systems \cite{Bloch08}. One important experimental demonstration is the paradigmatic Bose-Hubbard (BH) model~\cite{greiner02a}, in which bosonic atoms undergo the superfluid-Mott insulator quantum phase transition. This quantum phase transition is driven by the competition of atomic tunneling between lattice sites and short-range (on-site) two-body interactions. Recently, there is a growing interest in the study of extended BH models, in which the two-body interaction is long-range in the sense that it extends over several lattice sites. Extended BH models exhibit a host of quantum phases, amongst them supersolid and checkerboard phases~\cite{goral02,trefzger11}. Many studies employ dipolar atoms~\cite{griesmaier05} or polar molecules~\cite{ni08} to establish many-body quantum systems with long-range interactions. A review about recent progress along this direction can be found in Ref.~\cite{trefzger11}.

Extraordinarily strong and long-range multipolar interactions are also present between atoms excited to Rydberg states. Here the interaction strength can exceed that between groundstate atoms by more than ten orders of magnitude. The coherent excitation of Rydberg atoms by lasers has recently been demonstrated in a series of experiments (for a review see~\cite{saffman10} and references therein) and many theoretical and experimental groups have studied the strongly correlated many-body dynamics of Rydberg gases~\cite{heidemann07,weimer08,reetz08,pohl10,garttner12} and Rydberg atoms confined to optical lattices~\cite{viteau11,olmos09,weimer10,sela11,ji11,lesanovsky11,mukherjee11,mayle11,lauer12}.

The long-range interaction between Rydberg atoms is usually not of direct use in the context of extended BH models as the atomic motion due to coherent tunneling between the sites usually takes place on a time scale that is a hundred to a thousand times longer than the lifetime of the Rydberg states. This limitation can be overcome by using Rydberg-dressing, i.e. a weak admixture of a Rydberg state to the electonic groundstate, rather than directly exciting Rydberg states resonantly. Such Rydberg-dressed groundstate atoms exhibit long-range interactions that are comparable with typical ultracold energyscales~\cite{honer10,henkel10,pupillo10}. A number of recent studies have addressed the dynamics in Rydberg-dressed atomic gases, such as the formation of supersolids~\cite{pupillo10,cinti10,henkel10,henkel11}, solitons~\cite{maucher11} and collective many-body excitations~\cite{honer10} as well as excitation transport~\cite{wuester11}.
\begin{figure}
\begin{center}
\includegraphics[width=0.95\columnwidth]{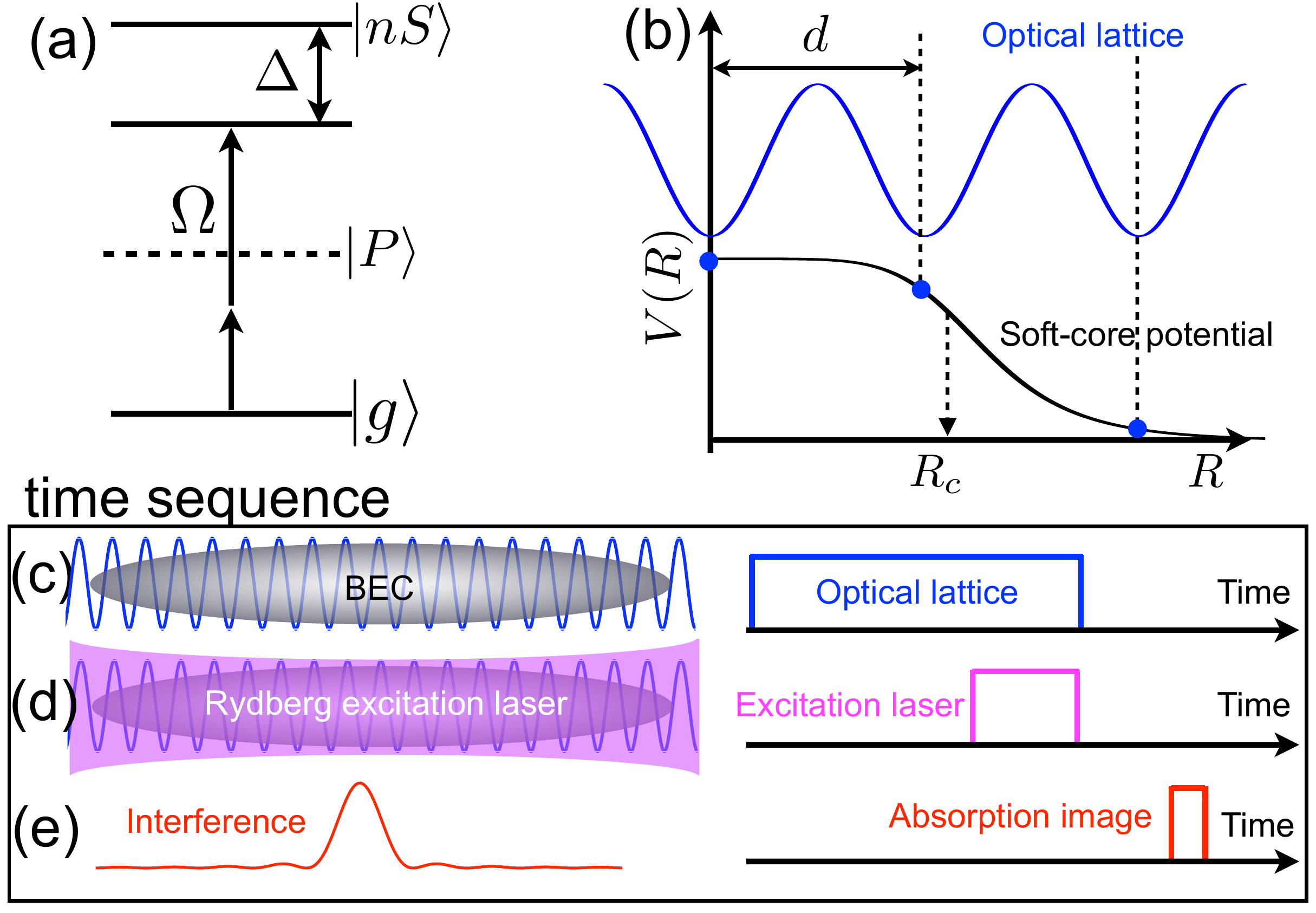}
\end{center}
\caption{(a) Schematics of the two-photon coupling between groundstate and Rydberg state. The laser which couples the state $|g\rangle$ to the intermediate $|P\rangle$-state is far detuned from resonance such that the groundstate $|g\rangle$ and the Rydberg $|nS\rangle$ state form an effective two-level system. Here $\Omega$ is the effective Rabi frequency and $\Delta$ the (two-photon) detuning of the excitation laser frequencies with respect to the atomic transition. (b) Effective interaction potential between dressed atoms and its length scales in relation to the optical lattice. At short distances ($R\ll R_\mathrm{c}$) the interaction potential is constant and at large distances ($R\gg R_\mathrm{c}$) it is of van-der Waals type. (c,d,e) Envisioned experimental sequence and timings. Firstly (c), a BEC is prepared in an optical lattice. Secondly (d), an off-resonant laser coupling the groundstate to the Rydberg state is applied for a certain time. After the excitation laser pulse, the optical lattice is switched off immediately and the dressed groundstate atoms expand freely. Finally (e), interference patterns are recorded by taking absorption images of the expanded atomic cloud.}
\label{fig:level}
\end{figure}

Here we consider the far off-resonant laser dressing of the electronic groundstate with a Rydberg $nS$ state, where $n$ is the principal quantum number and $S$ corresponds to the electronic angular momentum quantum number $l=0$. The laser excitation is achieved by a two-photon process, as illustrated in Fig.~\ref{fig:level}a. In the regime where the effective Rabi frequency $\Omega$ is much smaller than the detuning $\Delta$ the laser admixes merely a fraction of the Rydberg $nS$ state to the groundstate. As shown in Refs. \cite{henkel10,honer10,pupillo10} this induces an effective interaction potential between atoms which is sketched in Fig.~\ref{fig:level}b. The hallmark of this potential is the characteristic change of its behavior in the vicinity of the distance $R_\mathrm{c}$, which - depending on laser detuning and the Rydberg state - typically ranges from $1\mu$m to a few $\mu$m: At large inter-atomic distance $R\gg R_\mathrm{c}$, the potential decreases proportional to $1/R^6$, originating from the van der Waals (vdW) interaction between Rydberg atoms in $nS$ state. At short distances $R\ll R_\mathrm{c}$, the interaction potential levels off to a constant value. This soft-core behavior is a consequence of the Rydberg blockade effect~\cite{lukin01}, which inhibits the simultaneous excitation of nearby atoms to Rydberg states.

In this work we will investigate the effect of the Rydberg induced interaction on a lattice gas of groundstate atoms that is prepared in a superfluid state. In particular we are interested in studying the collapse and revival dynamics of the interference pattern that emerges upon the release of the atoms from the lattice. It has been demonstrated that collapses and revivals of a superfluid state released from an optical lattice provide important information on two-body interactions~\cite{greiner02,will10}. For example, complete and periodic revivals are expected if atoms interact through short-range interactions~\cite{greiner02}. In this work, we find that the peculiar shape of the interaction potential between Rydberg-dressed atoms gives rise to characteristic features in the collapse and revival dynamics of the interference pattern which originate from the competition of short-range and long-range parts of the interaction. Interference experiments therefore provide a way to probe the presence and to characterize the effective interaction between Rydberg-dressed atoms.

The paper is organized as follows. In Sec.~\ref{sec:hamiltonian}, we derive an effective extended BH model that governs the dynamics of Rydberg dressed atoms in an optical lattice. In Sec.~\ref{sec:wavefunction} we study the evolution of a superfluid state under the influence of the dressing, using the superfluid order parameter. In Sec.~\ref{sec:interference} we analyze the interference pattern of dressed atoms released from a one- and two-dimensional optical lattice. We conclude and provide experimental parameters in Sec.~\ref{sec:conclusion}.

\section{Effective Hamiltonian}
\label{sec:hamiltonian}
The system we consider here consists of $N$ bosonic atoms distributed over $L^D$ lattice sites, where $D$ is the spatial dimension. In our treatment, the Hamiltonian of the dressed groundstate atoms is obtained in two steps. First, we calculate the Born-Oppenheimer many-body interaction potential by a perturbative diagonalization of the electronic Hamiltonian. With this interaction potential, we then derive an extended BH Hamiltonian that governs the external dynamics of the dressed groundstate atoms in the optical lattice.

The internal level structure of the atoms is modeled by two states, the electronic groundstate $|g\rangle$ and the Rydberg $|nS\rangle$ state. These two electronic states are coupled by a two-photon transition with an effective Rabi frequency $\Omega$ as shown in Fig.~\ref{fig:level}a. The Hamiltonian for the internal (electronic) degrees of freedom of the atomic ensemble is given by (using $\hbar=1$ and the rotating-wave-approximation)
\begin{equation*}
H_\mathrm{e}=\sum_jH_j+\frac{1}{2}\sum_{j\neq k}V_{\rm{vdW}}({\bf r}_j,{\bf r}_k)|nS\rangle_j\langle nS|\otimes |nS\rangle_k\langle nS|
\end{equation*}
where $H_j=\Omega(|g\rangle_j\langle nS|+\mathrm{h.c.})+\Delta|nS\rangle_j\langle nS|$ is the single atom Hamiltonian. Here  $V_{\rm{vdW}}({\bf r}_j,{\bf r}_k)=C_6/|{\bf r}_j-{\bf r}_k|^6$ is the vdW interaction that is present between two Rydberg atoms at positions ${\bf r}_j$ and ${\bf r}_k$. Typical values for the dispersion coefficient $C_6$ in the case of rubidium-87 can be found in \cite{singer05, olmos11}. We do not consider the two-body contact interaction between groundstate atoms. This interaction can be switched off for instance by a Feshbach resonance~\cite{chin10}.

In our system, the electronic dynamics (Rydberg excitation and van-der-Waals interaction) takes place on a timescale ($\sim\, \mu$s) that is orders of magnitude faster than that of the external motion of ultracold atoms ($\sim\,$ ms). The huge difference in these time scales permits the use of the Born-Oppenheimer approximation for treating the electronic parts of the Hamiltonian. Rydberg dressing implies a far off-resonant excitation, i.e., $|\Omega/\Delta|\ll 1$ which allows to adiabatically eliminate the atomic Rydberg states. Using $\Delta>0$ we obtain in fourth order perturbation theory (in the small parameter $\Omega/\Delta$) a Born-Oppenheimer surface which can be written as a constant plus a sum of two-body interactions~\cite{henkel10,pupillo10},
\begin{equation}
\label{eq:born}
V_{\rm{BO}}=N\left(\frac{\Omega^4}{\Delta^3}-\frac{\Omega^2}{\Delta}\right)+\frac{1}{2}\sum_{j\neq k}V({\bf r}_j,{\bf r}_k)
\end{equation}
with
\begin{eqnarray*}
  V({\bf r}_j,{\bf r}_k)=\frac{g\,R_\mathrm{c}^6}{|{\bf r}_j-{\bf r}_k|^6+R_\mathrm{c}^6}.
\end{eqnarray*}
Here $g=2\Omega^4/\Delta^3$ and $R_\mathrm{c}=(C_6/2\Delta)^{1/6}$ represents the characteristic length scale of the soft-core interaction potential sketched in Fig.~\ref{fig:level}b. In the following, we will neglect the constant terms in Eq. (\ref{eq:born}) which correspond to the second and fourth order light shift.

Let us now turn to the discussion of the external (motional) degrees of freedom. The Rydberg dressed atoms are trapped in a $D$-dimensional optical lattice potential $V_\mathrm{L}({\bf r})$ (lattice spacing $d$) which is experimentally created by an optical standing wave~\cite{greiner02}. For $D<3$ we assume that there is a tight transverse confining potential with respect to which the atoms are in the motional groundstate. We furthermore neglect differential dephasing of the electronic groundstate with respect to Rydberg state which would be caused by different effective trapping frequencies due to varying AC-polarizabilities of the two states. Such dephasing of the motional degrees of freedom can be minimized by choosing a particular frequency of the standing wave light field that produces the optical potential in conjunction with an appropriate choice of the Rydberg state~\cite{mukherjee11}. With these approximations the second quantized Hamiltonian for the external motion of the dressed atoms reads
\begin{eqnarray}
\label{eq:2ndhamiltonian}
H &=&\int d{\bf r} \Psi^{\dagger}({\bf r})[-\frac{\hbar^2}{2m}\nabla^2+V_\mathrm{L}({\bf r})]\Psi({\bf r}) \nonumber \\ &+& \frac{1}{2}\int d{\bf r} d{\bf r}'\Psi^{\dagger}({\bf r})\Psi^{\dagger}({\bf r}')V({\bf r},{\bf r}')\Psi({\bf r}')\Psi({\bf r})
\end{eqnarray}
where $\Psi({\bf r})$ is the bosonic field operator of the dressed atoms. We expand the field operator $\Psi({\bf r})$ in terms of Wannier states $w({\bf r})$ in the lowest Bloch band of $V_\mathrm{L}({\bf r})$ such that $\Psi({\bf r})=\sum_jb_j w({\bf r}-{\bf r}_j)$ where $b_j$ is the annihilation operator of a dressed atom at site $j$ and ${\bf r}_j$ is the spatial position around which the Wannier states are centered. We consider the case of a deep optical lattice where these Wannier states are strongly localized on a single lattice site, i.e. the variance of the Wannier wave function is much smaller than the lattice spacing $d$. Here the Hamiltonian becomes that of an extended BH model
\begin{equation}
\label{eq:fullhamiltonian}
H=-J\sum_{\langle j,k\rangle}b^{\dagger}_{j}b_{k}+\frac{g}{2}\sum_{j}n_{j}(n_{j}-1)+\frac{1}{2}\sum_{\langle {j},{k} \rangle}\gamma(j-k)n_{j}n_{k},
\end{equation}
with $\langle j,k\rangle$ indicating the summation over neighboring lattice sites and $n_j=b^{\dagger}_{j}b_{j}$. The constants $J$ and $g$ refer to the tunneling matrix element and the on-site short-range interaction, respectively, while
\begin{eqnarray*}
  \gamma(m)=\frac{g\,R_\mathrm{c}^6}{|m|^6d^6+R_\mathrm{c}^6}
\end{eqnarray*}
characterizes the long-range interaction among atoms located in distant sites $j$ and $k$. Note that in general the index $j$ is a vector for $D>1$. For example, $j=(j_x,j_y,j_z)$ in a three-dimensional optical lattice.

\section{Collapse and revival of the superfluid order parameter}
\label{sec:wavefunction}
In this section, we study the effect that the Rydberg dressing has on the coherence properties of the atoms. The situation we have in mind is depicted in Fig.~\ref{fig:level}c-e. At first (Fig.~\ref{fig:level}c) one prepares - in the absence of the dressing laser - a superfluid state which is the groundstate of Hamiltonian (\ref{eq:fullhamiltonian}) with $g=0$. In the second step (Fig. \ref{fig:level}d) one applies the dressing laser for a time $t$. Here we consider a regime in which the parameters are chosen such that $|J|\ll |g|, |\gamma(m)|$, i.e. the atomic tunneling can be neglected during the time the dressing laser is switched on. The system then evolves approximately under the Hamiltonian (\ref{eq:fullhamiltonian}) with $J=0$. We are interested in the coherence properties of the superfluid state after this procedure. To this end we will study the superfluid order parameter in this section, while in the next section the probing of the coherence properties via an interference experiment (Fig. \ref{fig:level}e) is discussed.

The initial superfluid state of the atoms is given in terms of many-body Fock states,
\begin{equation}
 |{\rm{SF}}\rangle=\frac{1}{\sqrt{N!}}\left[\frac{1}{\sqrt{L}}\sum_{j}^Lb_{j}^{\dagger}\right]^N|0,0,\dots\rangle
 \label{eq:sfs}
 \end{equation}
where $|0,0,\cdots\rangle$ is the product of vacuum states of sites $j=1,2,\dots$. We work in the limit of a large atom number $N$ where we can replace the exact superfluid state by
\begin{equation}
|{\rm{SF}}\rangle\approx\prod_{j}|\alpha_{j}\rangle
\label{eq:sfc}
\end{equation}
with $|\alpha_{j}\rangle$ being a coherent state of atoms in a single lattice site defined as,
 \begin{equation}
 |\alpha_{j}\rangle={\text{e}}^{-\frac{|\alpha_{j}|^2}{2}}\sum_{n=0}^{\infty}\frac{\alpha_{j}^n}{\sqrt{n!}}|n\rangle.
 \end{equation}
We will use this representation as initial state for the following calculations and will moreover exclusively consider a homogeneous system where the average atom number in each lattice site is constant, i.e., $\alpha_{j}=\alpha$ with $|\alpha|^2=N/L$.

When the Rydberg excitation laser is applied, the superfluid state evolves according to $|\text{SF}(t)\rangle=\exp[-iH t]|\text{SF}(t=0)\rangle$. We will now discuss the effect of the dressing on the superfluid order parameter defined as
\begin{eqnarray*}
  \phi^{(D)}(t)=\langle\text{SF}(t)|b_{j}|\text{SF}(t)\rangle=\langle b_j\rangle=\alpha\exp\left[|\alpha|^2F_{D}(t)\right].
\end{eqnarray*}
We consider $D$-dimensional optical lattices with linear ($D=1$), square ($D=2$) and cubic ($D=3$) geometry. We will focus our analysis on the "phase factor" $F_D(t)$ as it encodes the full dynamics of the order parameter $\phi^{(D)}(t)$ and has the advantage that it is independent of the particle density. For the following discussion we will furthermore consider a parameter regime in which the interactions between distant atoms, i.e. atoms whose position labels obey $|j-k|>2$, is negligible. This is the situation depicted in Fig. \ref{fig:level}b. However, our results can straightforwardly be generalized. Let us now study $F_D(t)$ for lattices with different spatial dimension $D$:

\noindent (i) {\it  One-dimensional lattice.} Taking into account the on-site, nearest-neighbor $\gamma(1)$ and next-nearest-neighbor $\gamma(2)$ interaction we find
\begin{eqnarray}
F_1(t)&=&-[5-\text{e}^{-igt}-2\text{e}^{-i\gamma(1) t}-2\text{e}^{-i\gamma(2) t}].
\label{eq:matterwave}
\end{eqnarray}
In general we expect the two-body interactions to result in a collapse and revival of the superfluid order~\cite{imamoglu97,greiner02}. The collapse dynamics can be obtained by analyzing the real part of $F_1(t)$ in the vicinity of $t=0$. A Taylor expansion up to second order in $t$ yields
\begin{eqnarray}
\text{Re}[F_1(t)]\approx -\left[\frac{g^2}{2}+ \gamma^2(1)+ \gamma^2(2)\right]t^2=-\frac{1}{2}\left(\frac{t}{\tau_1}\right)^2
\label{eq:ft}
\end{eqnarray}
where $\tau_1$ is the collapse time, given as
\begin{eqnarray*}
   \tau_1\approx \frac{1}{\sqrt{g^2+2\gamma^2(1)+2\gamma^2(2)}}.
\end{eqnarray*}
This expression clearly indicates that the collapse is faster the stronger the two-body interaction. The possibility of revivals of the initial state is strongly determined by the long-range part of the interaction. Complete revivals occur when all phase factors appearing in exponential terms of $F_1(t)$ are simultaneously multiples of $2\pi$. As the ratios between the various possible interaction strengths are in general irrational, complete revivals are unlikely and one can rather expect partial revivals at a finite time. Before discussing this aspect in more detail let us briefly provide the results for two- and three-dimensional lattices.

\noindent (ii) {\it Two-dimensional lattice.} In two dimensions the truncation of the interaction to atoms whose indices obey $|j-k|\leq2$ means that we have to take into account at most next-next-nearest neighbor interactions. This yields
\begin{eqnarray*}
F_2(t)=-[13-\text{e}^{-igt}-4\text{e}^{-i\gamma(1) t}
- 4\text{e}^{-i\gamma(\sqrt{2}) t}-4\text{e}^{-i\gamma(2) t}]
\end{eqnarray*}
and a collapse time
\begin{eqnarray*}
\tau_2\approx \frac{1}{\sqrt{g^2+4\gamma^2(1)+4\gamma^2(\sqrt{2})+4\gamma^2(2)}}.
\end{eqnarray*}

\noindent (iii) {\it Three-dimensional lattice.} In this case one obtains
\begin{eqnarray*}
F_3(t)&=&-[33-\text{e}^{-igt}-6\text{e}^{-i\gamma(1) t}-12\text{e}^{-i\gamma(\sqrt{2}) t}\nonumber\\
&&-8\text{e}^{-i\gamma(\sqrt{3})t}-6\text{e}^{-i\gamma(2) t}],
\end{eqnarray*}
with a corresponding collapse time
\begin{eqnarray*}
\tau_3\approx \frac{1}{\sqrt{g^2+6\gamma^2(1)+12\gamma^2(\sqrt{2})+8\gamma^2(\sqrt{3})+6\gamma^2(2)}}.
\end{eqnarray*}

\noindent Generally the collapse time becomes shorter for higher dimensional systems, since the number of contributing interaction terms grows with increasing dimensions (as a consequence of the increasing coordination number). For the purpose of illustration let us now analyze the case $D=2$ in more detail. In particular, we intend to study $F_2(t)$ as a function of the ratio between the lattice spacing $d$ and the length scale $R_\mathrm{c}$ which demarcates the soft core of the effective interaction potential among Rydberg-dressed atoms.
\begin{figure}
\begin{center}
\includegraphics[width=3.3in]{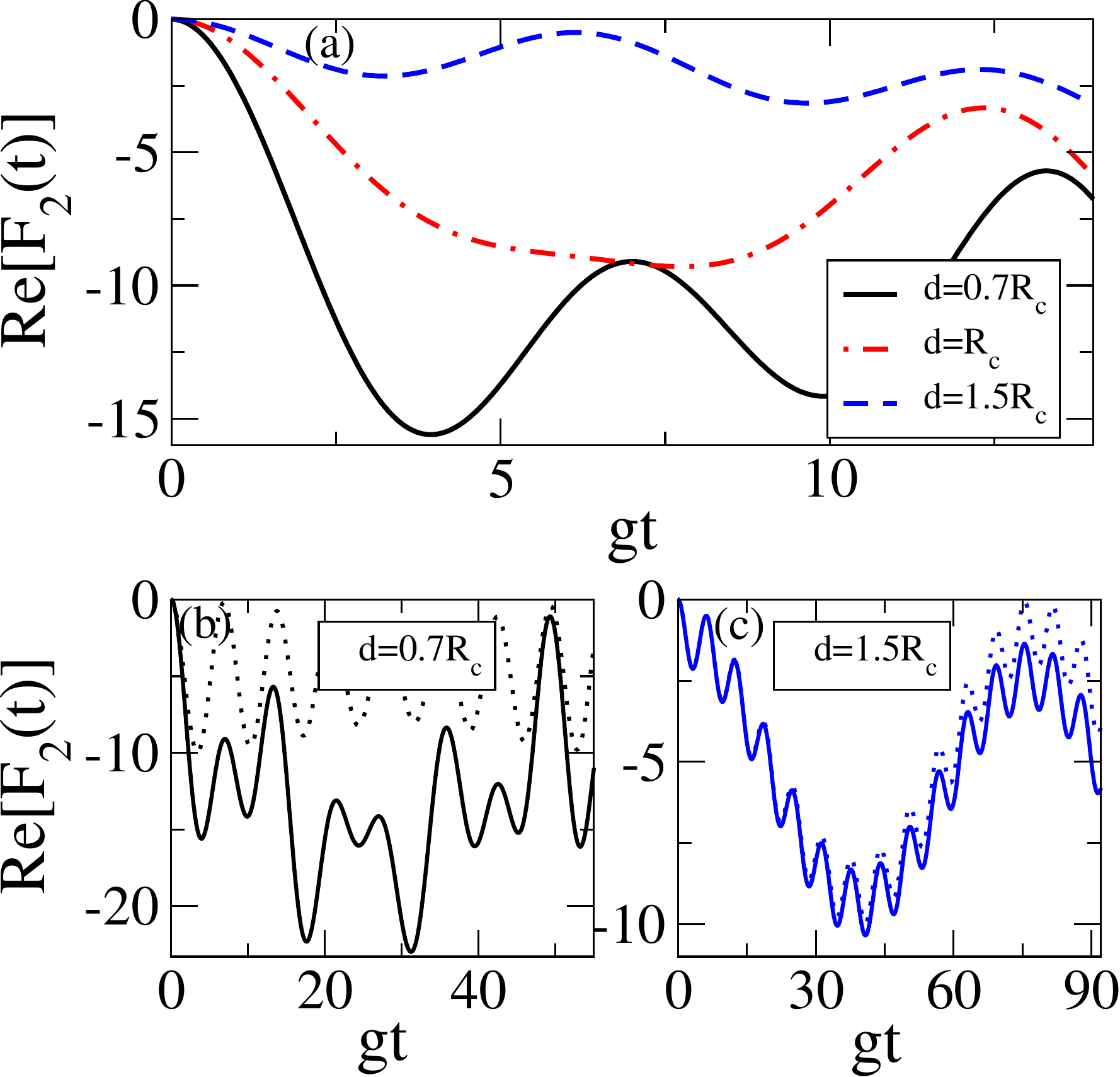}
\end{center}
\caption{(a) Dynamical evolution of $F_2(t)$ for various values of $d/R_\mathrm{c}$. (b, c) Dotted curves show calculations merely taking into account the on-site and nearest-neighbor interaction. Solid lines show calculations including all interactions up to next-next-nearest neighbors.}
\label{fig:matterwave}
\end{figure}
In Fig. \ref{fig:matterwave} we present data for $d/R_\mathrm{c}=\{0.7,1.0,1.5\}$. The smaller $d/R_\mathrm{c}$ the more terms of the long-range tail of the interaction actually contribute to the denominator of $\tau_2$. This leads to a more rapid collapse of the order parameter the smaller $d/R_\mathrm{c}$ becomes, which is clearly shown by the data. In fact, all dimensions have in common that for $d/R_\mathrm{c}>1$, the behavior of the function $F_D(t)$ is determined mainly by the on-site and nearest-neighbor interaction. Compared to those the other interaction terms are negligible which is owed to the fast drop of the soft-core interaction potential at large inter-site separations. To demonstrate this explicitly, we calculate $F_2(t)$ by neglecting interaction beyond nearest neighbors, i.e., $\gamma(m)=0$ for $m>1$. The results and a comparison with the full calculation are shown in Fig.~\ref{fig:matterwave}b and c. For $d/R_\mathrm{c}=0.7$ the two calculations deviate already beyond $gt=2$. However, for $d/R_\mathrm{c}=1.5$ agreement for much longer times (here until $gt=30$) is achieved. Thus, for sufficiently short times the contribution of the long-range interaction terms to the dynamical evolution of $F_2(t)$ is strongly suppressed.

\section{Interference pattern}
\label{sec:interference}
The dynamical evolution of the superfluid order can be made visible from an interference experiment as was shown in Ref. \cite{greiner02}. To create the interference pattern one simultaneously switches off the optical lattice and the Rydberg excitation laser. After a time-of-flight period, the interference pattern in the spatial density distribution (see Fig.~\ref{fig:level}e) is formed and made visible by taking an absorption image of the atomic cloud. This interference pattern is characterized by the quasi-momentum distribution~\cite{kashurnikov02,gerbier05}
\begin{eqnarray}
S({\bf {\tilde k}})&=& \sum_{{m}{n}}^L\text{e}^{i{\bf { \tilde{k}}}\cdot ({\bf r}_{m}-{\bf r}_{n})}\langle b_{m}^{\dagger}b_{n}\rangle.
 \end{eqnarray}
Here ${\bf \tilde{k}}$ is the quasi-momentum with components $\tilde{k}_{\xi}=\nu_\xi\times2\pi/(dL)$ where $\nu_\xi=1, 2, \cdots, L$ and, e.g., $\xi=x,y,z$ when $D=3$. For convenience we furthermore define the dimensionless quantity ${\bf k}=d\,{\bf \tilde{k}}$.

\begin{figure}
\begin{center}
\includegraphics[width=3.5in]{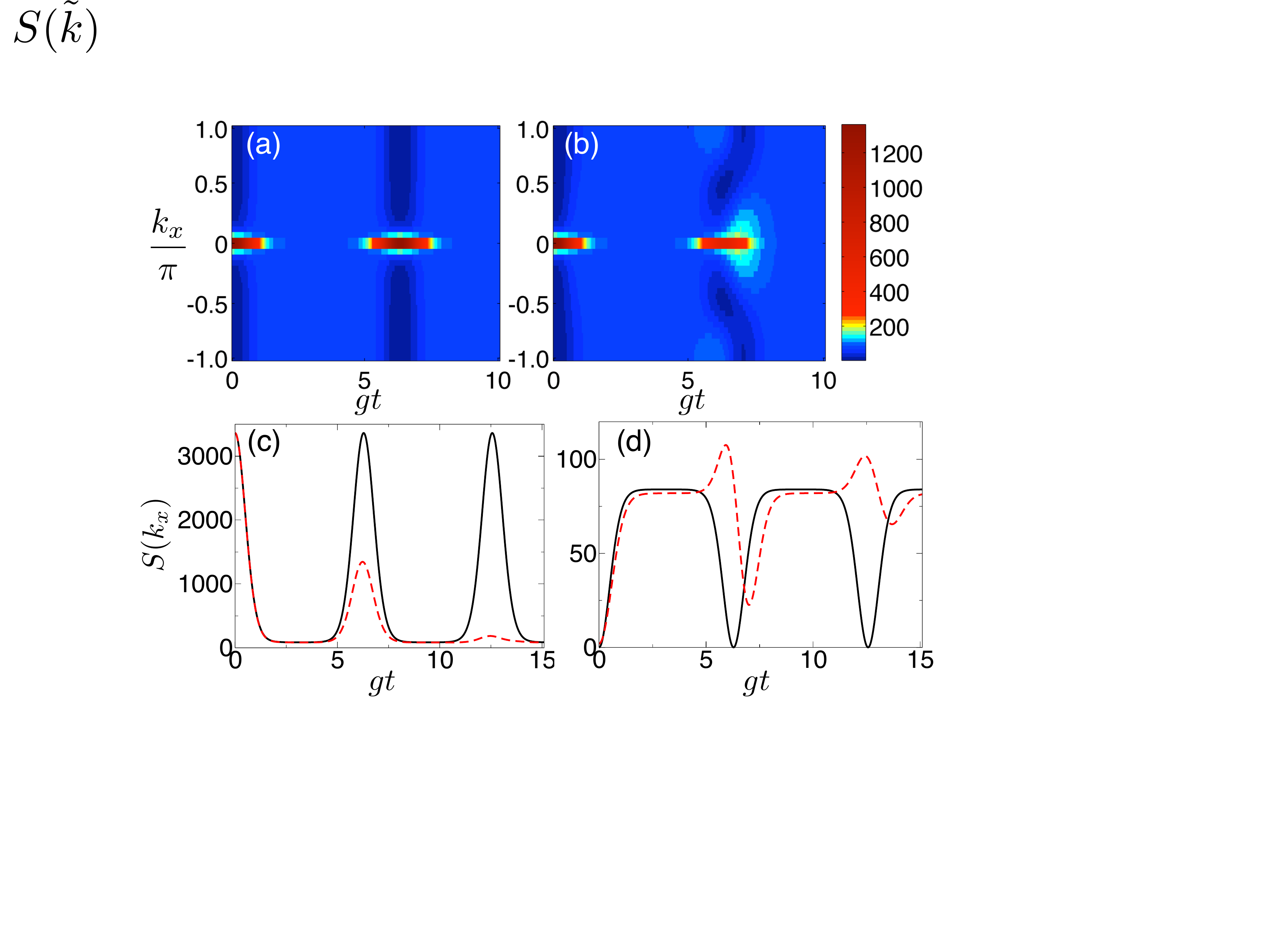}
\end{center}
\caption{Quasi-momentum distribution of $N=82$ atoms released from a one-dimensional optical lattice with $L=41$ sites. (a) On-site interaction only, i.e. $\gamma(m)=0$ (corresponding to $d/R_\mathrm{c}\gg 1$), (b) On-site and nearest-neighbor interaction for $d/R_\mathrm{c}=1.5$ ($\gamma(1) \approx 0.08g$).  (c,d) Cuts through $S({k_x})$ at different values of the quasi-momentum. The panels show cuts along ${k_x}=0$ (c) and ${k_x}=\pi$ (d). The solid/dashed curves correspond to the parameters of panel (a)/(b). }
\label{fig:int1d}
\end{figure}

Let us now study the effect of the Rydberg-dressing on $S({\bf k})$. We will consider here merely on-site and nearest-neighbor interactions of the extended BH model. This approximation is well justified provided that $d/R_\mathrm{c}>1$, as shown in the previous section. We furthermore will only focus on one- and two-dimensional lattices.
In order to calculate $S({\bf k})$, we first evaluate the matrix elements of the single particle density matrix, $\langle b_{j}^{\dagger}b_{k}\rangle=\langle {\rm SF}(t)|b_{j}^{\dagger}b_{k}|{\rm SF}(t)\rangle$ which are given in Appendix~\ref{app:correlation}. Using this data we obtain the following results:

\noindent (i) {\it One-dimensional lattice.} In order to be able to distinguish contributions of the on-site and nearest-neighbor interaction, we start with a situation where only the on-site interaction is present. This corresponds to a scenario that was experimentally studied in Ref. \cite{greiner02} and in case of the Rydberg-dressed atoms this is realized when $d/R_\mathrm{c}\gg 1$.  The quasi-momentum distribution $S(k_x)$ changes as a function of time as shown in Fig.~\ref{fig:int1d}a. Two main features can be identified. First, at $t=0$, $S({k_x})$ is centered around ${k_x}=0$. As the interaction time grows $S({k_x})$ collapses to flat distribution and at an even later time ($gt=2\pi$) it revives to its original value. The second feature is that the periodicity of $S(k_x)$ does not depend on $k_x$ (although the amplitude does). This is a direct consequence of the fact that the correlation function is factorizable, i.e., $\langle b^{\dagger}_{j}b_{k}\rangle=\langle b_{j}^{\dagger}\rangle\langle b_{k}\rangle$ for all $j$ and $k$ \cite{schachenmayer11}. The time-dependence thus becomes a global factor of $S({k_x})$.

This changes in the presence of the nearest-neighbor interaction. Due to a nonzero $\gamma(1)$, $\langle b^{\dagger}_{j}b_{k}\rangle$ is no longer factorizable but depends on the relative distance between respective sites (Appendix \ref{app:correlation}). This gives rise to strikingly different features in the quasi-momentum distribution. An example of this is shown in Fig.~\ref{fig:int1d}b for $d/R_\mathrm{c}=1.5$ where $\gamma(1) \approx 0.08g$. The noteworthy feature here is that the time-dependence of $S({k_x})$ depends on the quasi-momentum ${k_x}$. For example, at $gt=2\pi$, we find that $S({k_x})$ revives partially along $k_x=0$ (Fig.~\ref{fig:int1d}c), while along $k_x=\pi$ it shows a clearly different oscillation pattern (Fig.~\ref{fig:int1d}d).
\begin{figure}
\begin{center}
\includegraphics[width=3.4in]{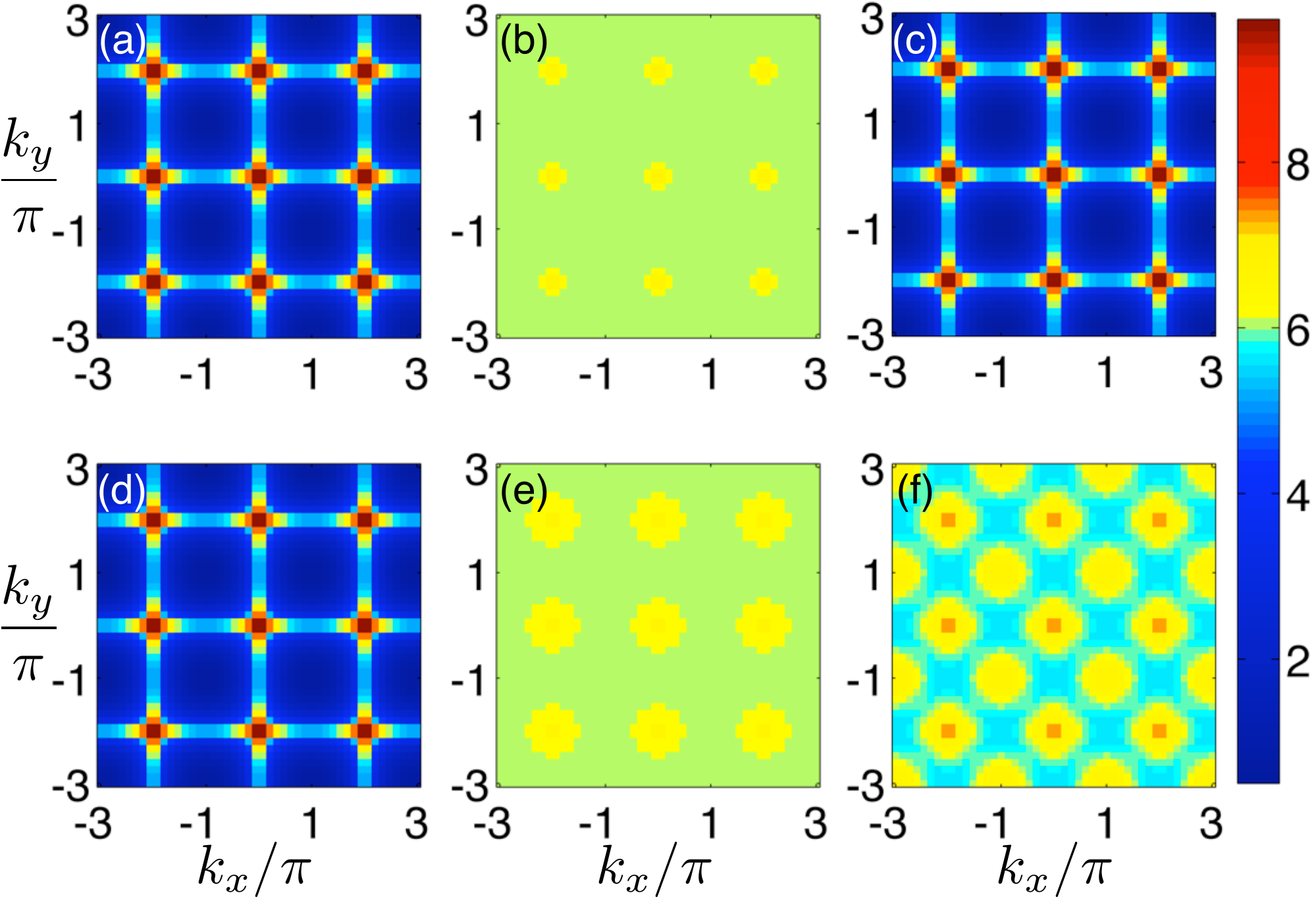}
\end{center}
\caption{Natural logarithm of the quasi-momentum distribution $S({\bf k})$ of atoms released from a two-dimensional $15\times 15$-lattice and an average particle number in each sites of $|\alpha|^2=2$. The figures show snapshots taken different times: $gt=0$ [(a) and (d)], $gt=\pi/2$ [(b) and (e)] and $gt=2\pi$ [(c) and (f)]. Panels (a)-(c) show data that has been obtained by taking only on-site interactions into account. The data shown in panels (d)-(f) includes also nearest-neighbor interactions.}
\label{fig:interference2d}
\end{figure}

\noindent (ii) {\it Two-dimensional lattice.} Let us now turn to the discussion of the interference pattern that emerges in the case in which the atoms are released from a two-dimensional optical lattice. In order to have a reference we again calculate at first the interference patterns for vanishing nearest-neighbor interaction. The main features in the interference pattern can be captured by analyzing time slices taken at $gt=0, \ \pi/2$ and $2\pi$. This data is shown in Fig.~\ref{fig:interference2d}a-c. Like in the one-dimensional case the time-evolution is periodic showing a first revival at $gt=2\pi$ and the expected periodicity \cite{greiner02}. In Fig.~\ref{fig:interference2d}d-f, we present the interference pattern that is obtained in case of a finite nearest-neighbor interaction ($d/R_\mathrm{c}=1.45$ and $\gamma(1)\approx 0.1g$) using the same time slices. We find that the height of the peaks of $S({\bf k})$ in Fig.~\ref{fig:interference2d}f is lowered compared to Fig.~\ref{fig:interference2d}c. The most striking feature is, however, the appearance of additional interference maxima in between the ones that occurred previously in the case of pure on-site interactions. These peaks occur because the nearest-neighbor interaction actually affects (indirectly) the phase relation between atoms that are separated by a distance $\sqrt{2}\,d$ (next-nearest neighbors) as shown in Appendix \ref{app:correlation}.

\section{Conclusions and Outlook}
\label{sec:conclusion}
In conclusion, we have shown that Rydberg-dressed atoms in an optical lattice give rise to an extended BH model. In particular, we have studied the dynamical evolution of an atomic superfluid under the influence of Rydberg dressing. The emerging long-range interactions result in a rapid collapse of the superfluid order parameter and in general allow only for partial revivals. Moreover, we have demonstrated that interference experiments can directly reveal the interaction between Rydberg-dressed atoms. The hallmark here is the emergence of additional interference maxima. In addition, the time-dependence of the quasi-momentum distribution is no longer a global factor. This leads to a time-dependence which strongly depends on value of the quasi-momentum.

We expect that our predictions can be probed in current experiments using Rydberg atoms in lattices, e.g. the one presented in Ref. \cite{viteau11} by Viteau {\it et al.} In this particular experiment Rydberg D-states of rubidium-87 with principal quantum numbers in the range $n=55-80$ are excited and the lattice spacing can be tuned within $0.42\ \mu \mathrm{m}<d<13\ \mu \mathrm{m}$. If one instead excites a Rydberg S-state with $n=60$ and uses a laser with the parameters $\Delta=2\pi\times 15$ MHz and $\Omega=2\pi\times 0.5$ MHz one obtains $R_\mathrm{c}\approx 4\ \mu \mathrm{m}$ as the characteristic length scale of the soft-core potential. This leaves sufficient freedom for tuning the ratio $d/R_\mathrm{c}$. The typical revival time of the interference signal evaluates to $t_0\approx 27 $ ms. This has to be compared with the effective lifetime of dressed groundstate atoms, $\tau_{\rm{eff}}=|\Delta/\Omega|^2\times \tau_{\rm{Ryd}}$ where $\tau_{\rm{Ryd}}$ is the bare lifetime of the Rydberg state. Using the above parameters, we obtain $\tau_{\rm{eff}}=225$ ms, which leads to a coherence time that in principle should permit the observation of the predicted dynamics.

\acknowledgments
We acknowledge C. Ates, B. Olmos and S. Genway for fruitful discussions. I. L. acknowledges support by EPSRC and through the Leverhulme Trust. W. L. acknowledges support by the EU (Marie Curie Fellowship).
\\ \\
\appendix
\section{Correlation functions}
\label{app:correlation}
Using the initial state Eq.~(\ref{eq:sfc}) and taking into account of on-site and nearest-neighbor interactions, $\langle b_{j}^{\dagger}b_{k}\rangle$ is calculated analytically. The matrix elements required for calculating the quasi-momentum distribution are presented below.

If $j=k$, we get $\langle b^{\dagger}_{j}b_{j}\rangle=|\alpha|^2$.

If $|j-k|=1$,
\begin{eqnarray}
\langle b_{j}^{\dagger}b_{k}\rangle &=&|\alpha|^2\text{e}^{-2|\alpha|^2[1-\cos(gt-\gamma(1)t)]}\text{e}^{-2(2D-1)|\alpha|^2[1-\cos\gamma(1)t]}. \nonumber
\end{eqnarray}

 If $|j-k|=\sqrt{2}$,
 \begin{equation}
\langle b_{j}^{\dagger}b_{k}\rangle=|\alpha|^2\text{e}^{-2|\alpha|^2[1-\cos gt]-4(D-1)|\alpha|^2[1-\cos\gamma(1)t]}.\nonumber
\end{equation}

 If $|j-k|=2$,
 \begin{equation}
\langle b_{j}^{\dagger}b_{k}\rangle=|\alpha|^2\text{e}^{-2|\alpha|^2[1-\cos gt]-2(2D-1)|\alpha|^2[1-\cos\gamma(1)t]}.\nonumber
\end{equation}

If $|j-k|>2$,
 \begin{equation}
\langle b_{j}^{\dagger}b_{k}\rangle=|\alpha|^2\text{e}^{-2|\alpha|^2[1-\cos gt]}\text{e}^{-4D|\alpha|^2[1-\cos \gamma(1)t]}.\nonumber
\end{equation}

\bibliography{LatticeRydberg}
\end{document}